# Lattice-mismatch Moiré laser with strong flatband coupling

Donghwee Kim[1]†, Chiwon Shin[1]†, Changi Kim[1], Gil-Woo Lee[2], You-Shin No[2], Jin-Kyu Yang[3], Heonsu Jeon[1]*, and Hong-Gyu Park[1]*

[1]Department of Physics and Astronomy, and Institute of Applied Physics, Seoul National University, Seoul 08826, Republic of Korea.

[2]Department of Physics, Konkuk University, Seoul 05029, Republic of Korea.

[3]Department of Optical Engineering, and Institute of Application and Fusion for Light, Kongju National University, Cheonan 31080, Republic of Korea.

*Corresponding authors. Email: hsjeon@snu.ac.kr and hgpark@snu.ac.kr

†These authors contributed equally to this work.

**Abstract**

Inter-cell and/or interlayer coupling in Moiré superlattices can generate flatbands and collective eigenmodes that enable emergent physical phenomena, motivating extensive exploration of Moiré-inspired photonic devices. However, the experimental validation of robust inter-cell interactions in Moiré photonic structures and the modulation of flatbands for specific photonic applications remain challenging. Here, we propose a lattice-mismatch Moiré cavity and demonstrate nanolasers enabled by strong flatband coupling. In contrast to a twist-angle Moiré cavity, a lattice-mismatch Moiré cavity provides a stable flatband frequency and a substantial enhancement in $Q$ factor compared to an isolated single-cell cavity, as the unit-cell size decreases. The photonic band-structure measurement of the small-unit-cell Moiré cavity by photoluminescence reveals pronounced flatbands. Cell-resolved spectroscopy further confirms the presence of flatbands by identifying resonant peaks that consistently emerge across unit cells in a Moiré cavity with a lattice mismatch of 102 nm, but not in a larger-unit-cell Moiré cavity with a mismatch of 60 nm. Furthermore, mode selection is achieved by reducing the center-hole size, thus isolating the hexapole mode from the degenerate dipole modes while maintaining strong inter-cell coupling. Consequently, we demonstrate a low-threshold hexapole flatband laser in a single mode. Therefore, the systematic modification of the relative lattice parameters of the two constituent lattices offers a promising strategy for developing Moiré nanolasers and flatband nanophotonic devices.

**INTRODUCTION**

Moiré superlattices, created by superimposing two periodic lattices with a slight misalignment, have received considerable attention due to the resultant superperiodicity, which can lead to emergent physical phenomena[1-11]. There are two types of Moiré patterns: twist-angle Moiré and lattice-mismatch Moiré[12,13]. The former originates from a relative rotation between lattices, while the latter occurs when lattices with different constants are overlaid. In condensed-matter Moiré systems, Moiré flatbands enhance the density of states and thus intensify electron-electron interactions[2-7]. For example, Moiré superlattices composed of two-dimensional materials including graphene and h-BN have been reported to host not only correlated phases such as superconductivity, but also interaction-driven topological phases, such as the integer and fractional quantum anomalous Hall effects[4-11]. Similarly, in photonics, same flatband features have been observed in Moiré-inspired structures, enhancing light-matter interactions and achieving nanocavities with superior quality ($Q$) factors and reduced mode volumes[13-31]. An adaptive sensor for hyperpolarimetric imaging has also been demonstrated using twist-angle Moiré structures[29]. In addition, cavity quantum electrodynamics has been shown by observing the Purcell effect between individual quantum dots and Moiré cavity modes[30,31].

Flatbands are fundamentally influenced by inter-cell coupling, leading to collective modes expected to extend across the entire Moiré superlattice[17-20]. However, experimental verification of such strong mode coupling in Moiré-based photonic structures is challenging and cannot be accomplished solely through band-structure measurements[19-21]. In addition, accurately adjusting a flatband to a specified frequency range and positioning it within a photonic bandgap are essential for the demonstration of robust nanophotonic devices[32]. Moreover, among the two primary methods of generating Moiré patterns, photonic applications have thus far been restricted to twist-angle Moiré configurations, whereas lattice-mismatch Moiré structures have yet to be investigated[16,18]. Consequently, the systematic modification of the relative lattice parameters of the two constituent lattices, including both the twist angle and the lattice-constant difference, is a crucial step in the development of more effective Moiré cavities[12,13,15,20,21].

In this work, we experimentally demonstrate strong flatband mode coupling and lasing action in a lattice-mismatch Moiré cavity. Systematic optical simulations reveal that, in contrast to twist-angle Moiré cavities, lattice-mismatch Moiré cavities exhibit reduced frequency sensitivity but a significant dependence of the $Q$ factor on the unit-cell size. In addition to band-structure measurements that agree well with the simulations, observation of the same mode at consistent wavelengths across multiple unit cells provides direct evidence of strong inter-cell mode coupling. Furthermore, by modifying the central hole, we achieve low-threshold single-mode lasing in a hexapole flatband mode, highlighting lattice-mismatch Moiré cavities as a viable platform for robust flatband nanolasers.

**RESULTS**

**Nontrivial lattice-mismatch Moiré phenomena**

While twist-angle and lattice-mismatch Moiré patterns often exhibit similar behavior in condensed-matter systems, they can produce different optical properties. Figures 1a and 1b compare twist-angle and lattice-mismatch Moiré cavity designs based on three-dimensional finite element method (FEM) simulations. In both cases, two triangular air-hole lattices are defined in a dielectric slab with a thickness of $h = 257$ nm and a hole diameter of $d = 212$ nm, and a Moiré superlattice is formed either by rotating the lattices by a relative twist angle $\delta\theta$ (Fig. 1a) or by introducing a lattice constant difference $\delta a$ (Fig. 1b). First, we selected three twist angles, $\delta\theta =$ 3.48°, 4.41°, and 6.01°, for the twist-angle Moiré cavity and calculated the flatband frequency and $Q$ factor near a wavelength of ~1.5 μm. These values of $\delta\theta$ correspond to commensurate twist angles that generate periodic Moiré supercells in triangular lattices. We then plotted the flatband frequency shift relative to the value at $\delta\theta = 3.48°$, $\Delta f$, and the $Q$ factor ratio with respect to an isolated single-cell cavity, $Q/Q_{single}$, as a function of $\delta\theta$ (Fig. 1a, right). The simulation results show that as $\delta\theta$ increases, $\Delta f$ increases up to 26.67 THz, while $Q/Q_{single}$ remains relatively constant with a maximum of ~26.72 (Supplementary Fig. 1).

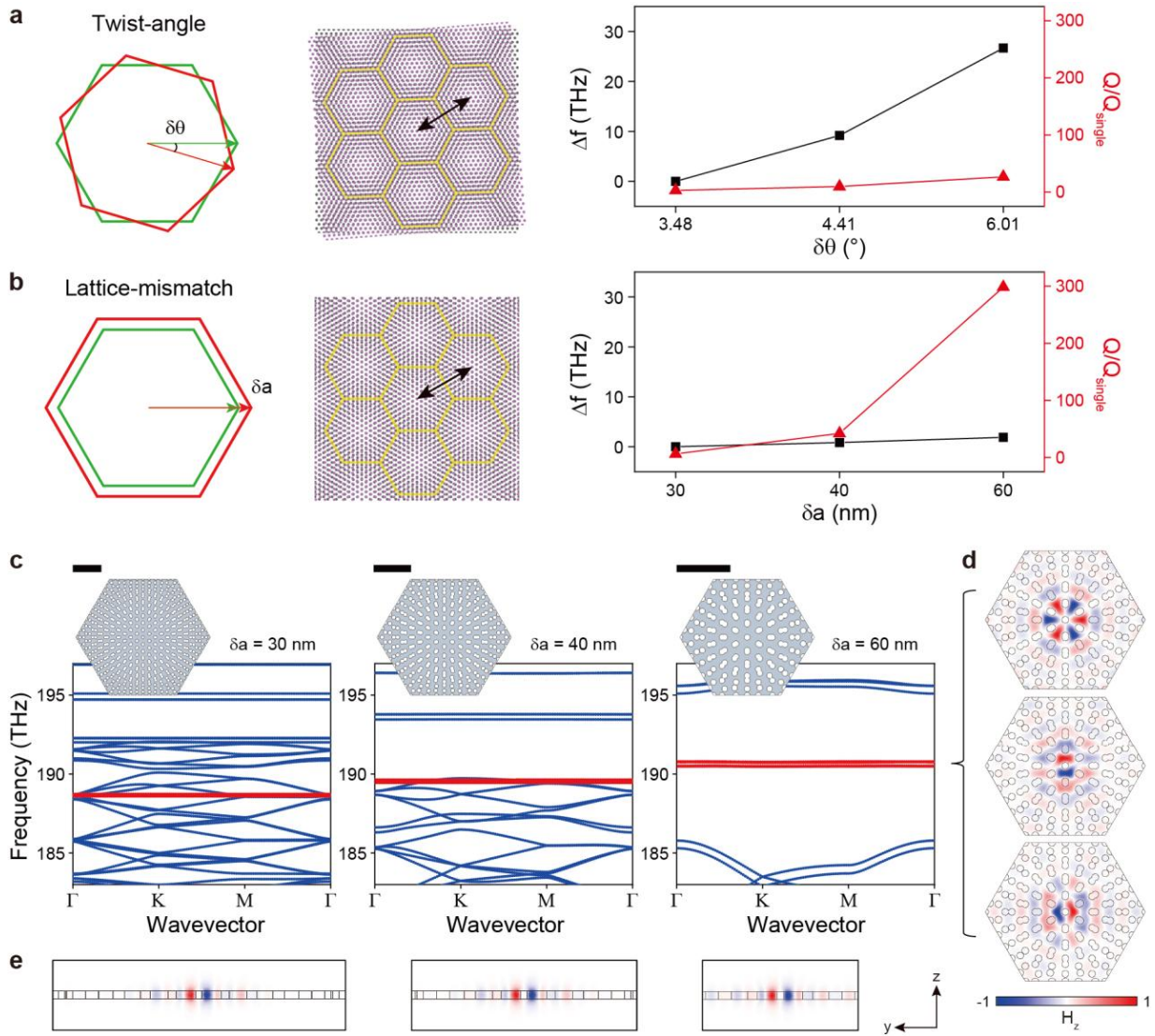

**Fig. 1. Twist-angle and lattice-mismatch Moiré cavity designs and their calculated properties. a-b,** Schematics (left), Moiré patterns (middle), and calculated flatband properties (right) for twist-angle (**a**) and lattice-mismatch (**b**) Moiré cavities. The Moiré patterns originate from the superposition of two triangular lattices characterized by a relative twist angle $\delta\theta$ (**a**) or a lattice-constant difference $\delta a = |a_1 - a_2|$ (**b**). The Moiré unit-cell size is indicated by a black arrow. Flatband calculations for a dielectric slab with refractive index of 3.33, thickness $h$ = 257 nm, and hole diameter $d$ = 212 nm show the frequency shift relative to the initial values ($\delta\theta$ = 3.48° for **a**; $\delta a$ = 30 nm for **b**). The results, $\Delta f$ (black squares; left y-axis) and the $Q$ factor ratio relative to an isolated single-cell cavity, $Q/Q_{single}$ (red triangles; right y-axis), are plotted as functions of $\delta\theta$ (**a**) or $\delta a$ (**b**). **c,** Calculated band diagrams of lattice-mismatch Moiré cavities for $\delta a$ = 30, 40, and 60 nm (left to right) within the frequency range of 183–197 THz, where the flatbands are highlighted in red. The corresponding Moiré unit-cell structures are shown as insets. Inset scale bars, 2 µm. **d,** Calculated $H_z$ field distributions of three flatband modes at $\delta a$ = 60 nm: the hexapole mode (top) and the doubly degenerate dipole modes ($p_x$ and $p_y$: middle and bottom). **e,** Calculated $H_z$ field distributions of the $p_y$ flatband mode at $\delta a$ = 30, 40, and 60 nm (left to right) on the $x$ = 0 plane, corresponding to the band diagrams in **c**.

Next, we perform the same simulations for lattice-mismatch Moiré cavities with $\delta a$ = 30, 40, and 60 nm, chosen such that the resulting Moiré unit-cell sizes correspondingly match those in the twist-angle configuration. The Moiré unit-cell sizes are 8.16 µm for ($\delta\theta$, $\delta a$) = (3.48°, 30 nm), 6.24 µm for (4.41°, 40 nm), and 4.32 µm for (6.01°, 60 nm), respectively (Supplementary Fig. 2). These unit-cell sizes in the lattice-mismatch Moiré cavities are defined by the lattice-constant ratio $a_1 : a_2$, corresponding to 8:9, 12:13, and 16:17, respectively (Supplementary Note 1). Because all other structural parameters were kept identical, the patterns depicted in Figs. 1a and 1b can be compared equitably. For example, the lattice-mismatch Moiré cavities exhibit nearly constant flatband frequencies, with $\Delta f$ varying by only 1.89 THz, together with a substantial enhancement in the $Q$ factor, reaching $Q/Q_{single}$ ~ 300. This behavior contrasts with that of twist-angle Moiré cavities, where $\Delta f$ varies with the unit-cell size while $Q/Q_{single}$ remains comparatively stable. Furthermore, for a given Moiré unit-cell size, the larger $Q$ enhancement in the lattice-mismatch case suggests that the flatband modes experience stronger inter-cell coupling and more effective suppression of radiative loss in the array compared to the twist-angle case.

Figure 1c shows the calculated band diagrams of lattice-mismatch Moiré cavities with $\delta a =$ 30, 40, and 60 nm in the frequency range of 183–197 THz, where the flatbands are highlighted in red. As $\delta a$ increases from 30 to 60 nm, the flatband evolves from overlapping with dispersive bands ($\delta a$ = 30 nm), to touching the band edge ($\delta a$ = 40 nm), and eventually to forming an in-gap flatband fully embedded within the photonic bandgap ($\delta a$ = 60 nm). The $H_z$ field profiles of three flatband modes, including a hexapole and two degenerate dipoles ($p_x$ and $p_y$), were also calculated for the lattice-mismatch Moiré cavity with $\delta a$ = 60 nm, which has a Moiré unit-cell size of 4.32 µm (Fig. 1d). The $H_z$ field distributions of the $p_y$ flatband mode on the $x = 0$ plane are shown below each band diagram (Fig. 1e). The origin of these three flatband modes can be understood using a simple tight-binding model (Supplementary Note 2).

The calculations reveal several key differences between the twist-angle and lattice-mismatch Moiré configurations. First, the band diagrams in Fig. 1c show that, in the lattice-mismatch Moiré cavities, the dispersive bands below the flatband appear as a pair of band sets that closely follow the parent bands of the constituent lattices[20,33], with variations in $\delta a$ shifting their relative alignment. This behavior contrasts with that of twist-angle Moiré structures, where the parent bands merge into a singular rebuilt band manifold. Second, despite variations in the Moiré unit-cell size, the flatband frequency in lattice-mismatch Moiré remains consistent, in contrast to the twist-angle one. This robustness is advantageous for cavity design. Third, in the small-unit-cell regime, lattice-mismatch Moiré cavities exhibit a pronounced $Q$-factor enhancement upon forming an array: whereas an isolated single-cell cavity has a low $Q$, the $Q$ increases markedly in the Moiré array. In contrast, the corresponding $Q$ enhancement in twist-angle Moiré is more than an order of magnitude smaller (Supplementary Fig. 2). This difference arises because, in lattice-mismatch Moiré cavities, increasing $\delta a$ makes the mode swell and its field distribution extend to the unit-cell boundary (Fig. 1e), therefore enhancing inter-cell mode coupling and resulting in the formation of collective modes that span the entire Moiré superlattice.

**Photonic band characterization and Moiré mode coupling**

To verify the presence of a flatband in the lattice-mismatch Moiré photonic structure, we performed band-structure measurements in a cavity with a Moiré unit-cell size of 2.40 μm (Fig. 2). This lattice-mismatch Moiré cavity was formed by superimposing two triangular lattices with lattice constants of 480 nm and 600 nm ($\delta a$ = 120 nm) and a hole diameter of $d$ = 216 nm. Notably, this structure was designed to have a reduced Moiré unit-cell size compared with the structures analyzed in the calculations of Fig. 1. Given the limited sample size, it is essential to measure the band structure using an array containing the maximum feasible number of unit cells. Reducing the unit-cell size increases the number of unit cells within the array, thereby enlarging the Brillouin zone in reciprocal space and improving the accuracy of the band-structure measurements.

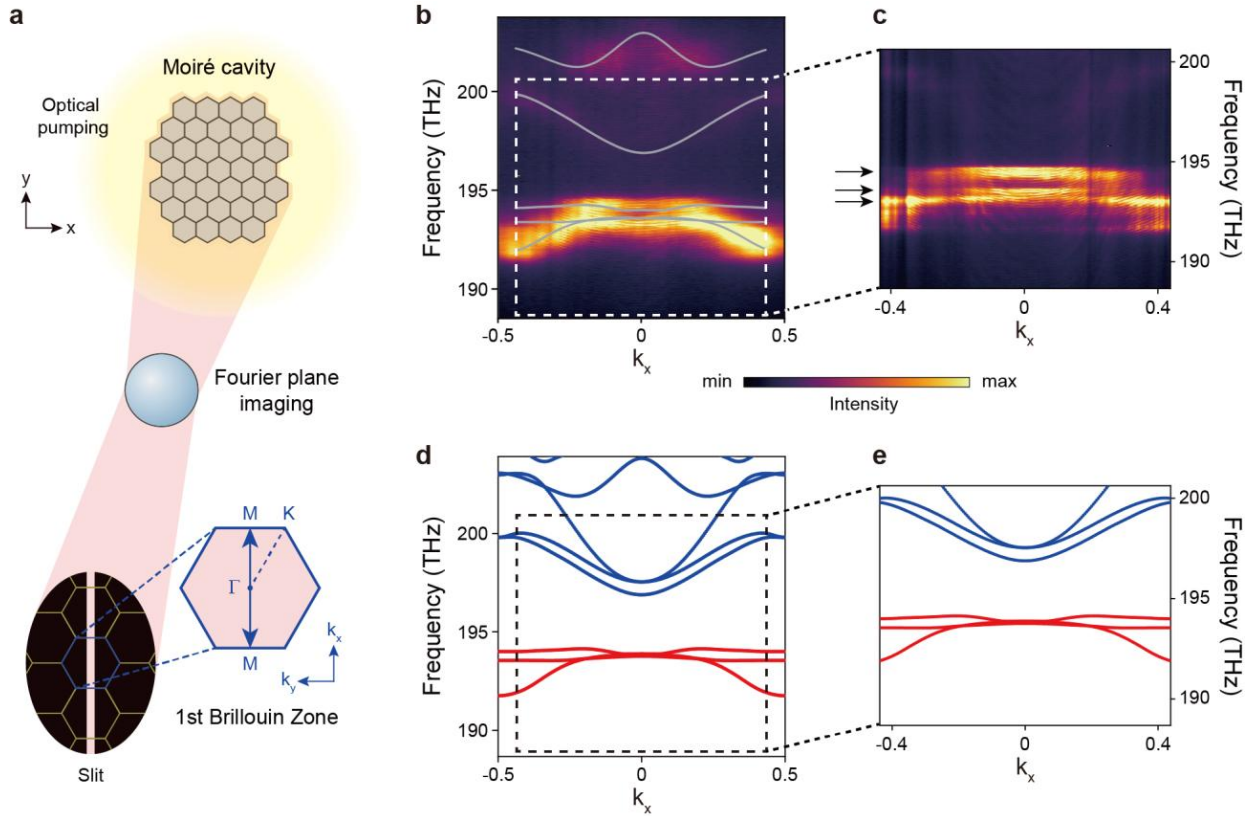

**Fig. 2. Photonic band measurement of a lattice-mismatch Moiré cavity. a,** Schematic of the band measurement setup. A more detailed setup is shown in Supplementary Fig. 3a. **b,** Measured band diagram of a lattice-mismatch Moiré cavity with a unit-cell size of 2.40 μm, acquired using a ×100 objective lens (N.A. = 0.85). Gray guidelines are overlaid for clarity. **c,** Measured band diagram of the same Moiré cavity acquired using a ×50 objective lens (N.A. = 0.42), corresponding to the frequency–$k_x$ window indicated by the dashed rectangle in **b**. Three black arrows indicate the flatbands. **d,** Calculated band diagram of the lattice-mismatch Moiré cavity with structural parameters identical to those in **b** and **c**. Flatbands are highlighted in red. **e,** Magnified view of the region indicated by the dashed rectangle in **d**, shown over the same $k_x$ range as in **c**.

Figure 2a shows a schematic of the band measurement setup using photoluminescence (PL) emission from the sample. In our experiment, the PL was collected at the Fourier plane, spatially filtered by a vertical slit, and spectrally dispersed by a spectrometer (see Methods and Supplementary Fig. 3a). The band diagram of the lattice-mismatch Moiré cavity was measured using a ×100 objective lens (N.A. = 0.85; Fig. 2b) and a ×50 objective lens (N.A. = 0.42; Fig. 2c), respectively. Reducing the numerical aperture (N.A.) from 0.85 to 0.42 provides a magnified view of the band diagram over a narrower $k_x$ range. Three flatbands were observed near $k_x = 0$ (Fig. 2b), and this feature is more clearly resolved in the magnified view (Fig. 2c). For comparison with the experiment, we calculated the band structure of the same cavity, and the results were plotted to match the corresponding measurement window (Fig. 2d). Figure 2e shows an enlarged view of the region (0.87 Brillouin zone) indicated by the dashed rectangle in Fig. 2d. The gray guidelines in Fig. 2b were derived from the calculated dispersion in Fig. 2d and overlaid for clarity.

The agreement between the measured and calculated band diagrams indicates that mode coupling in the lattice-mismatch Moiré cavity is well established. By implementing a reduced Moiré unit-cell size of 2.40 μm, we employed an array of 35 unit cells (Supplementary Fig. 3b) and obtained band diagrams from the central unit cell that closely match the calculated dispersion. The first Brillouin-zone size in this design is approximately 3.02 μm$^{-1}$, which lies within the measurable limits of both the ×100 and ×50 objective lenses, enabling a direct comparison between the measured and calculated $k$-space ranges. In particular, the three measured flatbands appear near 193.8 THz, in agreement with the calculations, and correspond to the highest-$Q$ flatband modes (hexapole, $p_x$, and $p_y$). Higher-frequency dispersive bands are also observed, and their frequency range is consistent with the calculated band diagram, supporting the validity of the band measurements. The comparatively weaker intensity of the lower-$Q$ dispersive bands provides additional evidence that the measured band features reflect the underlying mode properties. Away from $k_x = 0$, the bands exhibit a slight reduction in flatness; this effect is attributed to the reduced unit-cell size and is consistently observed in both experiment and simulation.

Next, we experimentally examined how the properties of lattice-mismatch Moiré cavities vary with the unit-cell size (Fig. 3). A key advantage of this lattice-mismatch Moiré approach is the straightforward tunability of the unit-cell size (Supplementary Note 1). We fabricated two lattice-mismatch Moiré cavities with unit-cell sizes of 3.06 μm (Figs. 3a-c) and 4.32 μm (Figs. 3d-f). The 3.06 μm unit cell was formed by superimposing two triangular lattices with lattice constants of 510 nm and 612 nm ($\delta a$ = 102 nm) and hole diameter $d$ = 255 nm. The 4.32 μm unit cell corresponds to the lattice-mismatch configuration with $\delta a$ = 60 nm in Fig. 1. By comparing the 4.32 μm unit cell with the smaller 3.06 μm unit cell (Supplementary Note 1), we quantitatively evaluate how the inter-cell coupling strength evolves with decreasing Moiré unit-cell size. Figures 3a and 3d show scanning electron microscopy (SEM) images of these two cavities. Each cavity contains seven core Moiré unit cells, numbered from 1 to 7, with a magnified view of a single Moiré unit cell shown in the top panel.

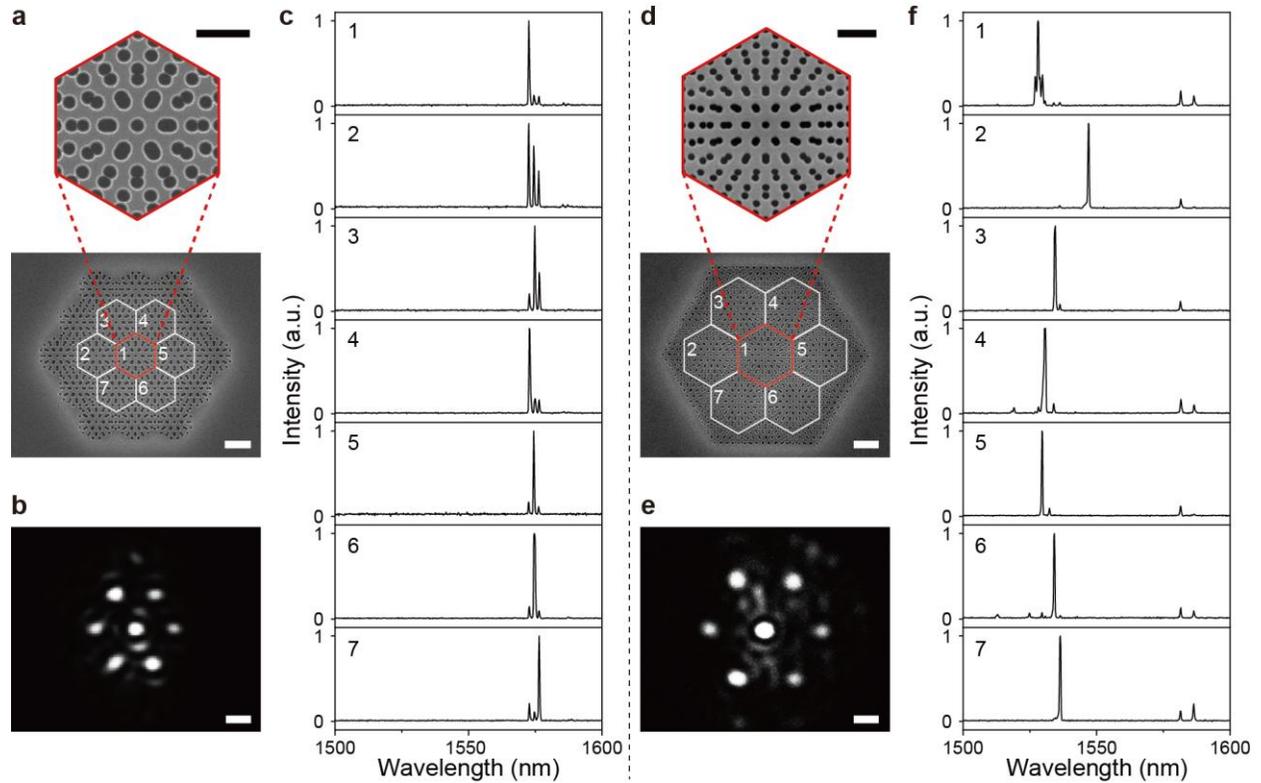

**Fig. 3. Optical characterization of lattice-mismatch Moiré cavities with varying unit-cell sizes.**
**a** and **d,** SEM images of lattice-mismatch Moiré cavities with unit-cell sizes of 3.06 μm (**a**) and 4.32 μm (**d**). Scale bars, 2 μm. Each cavity contains seven core Moiré unit cells, numbered from 1 to 7. Top, magnified view of a single unit cell. Scale bars, 1 μm. **b** and **e,** Measured real-space PL images from the cavities shown in **a** (**b**) and **d** (**e**). Scale bars, 2 μm. **c** and **f,** Measured PL spectra from the seven Moiré unit cells indicated in **a** (**c**) and **d** (**f**). Each panel shows the spectrum from the corresponding numbered unit cell within the wavelength range of 1500–1600 nm.

We then carried out room-temperature PL experiments on the fabricated cavities using pump laser pulses generated by a 980-nm laser diode with a repetition rate of 1 MHz and a duty cycle of 1%. The pump beam was focused onto the sample with a spot diameter of ~29.2 μm (see Methods). The resulting real-space PL images from the 3.06 μm and 4.32 μm cavities are shown in Figs. 3b and 3e, respectively. In both cases, bright emission spots appear at the centers of the seven Moiré unit cells, consistent with the calculated field confinement of the flatband modes (Fig. 1d). We also acquired PL spectra from each of the seven Moiré unit cells; the corresponding spectra in the telecommunication band are plotted in Figs. 3c and 3f, with each panel representing the spectrum from the Moiré unit cell of the same number. In the Moiré cavity with a unit-cell size of 3.06 μm, three emission peaks associated with the hexapole, $p_x$, and $p_y$ flatband modes were detected at 1572.8 nm, 1574.6 nm, and 1576.4 nm, respectively (Fig. 3c). These three peak wavelengths are consistent across all seven unit cells (cells 1–7). In contrast, in the cavity with a larger unit-cell size of 4.32 μm, the dominant emission peaks vary among unit cells are distributed over a broad wavelength range from 1478.0 nm to 1497.1 nm, with the spectral line shapes also differing between cells (Fig. 3f).

These observations indicate that, in lattice-mismatch Moiré cavities, the unit-cell size determines the extent of Moiré-induced mode coupling, which can be experimentally assessed through spectral alignment. Specifically, as the Moiré unit-cell size increases from 3.06 to 4.32 μm, the emission spectra measured from unit cells 1–7 exhibit pronounced cell-to-cell variations in peak positions, resulting in reduced spectral uniformity across the array. The weak mode coupling observed in the 4.32 μm cavity contrasts with the calculations presented in Fig. 1, suggesting that a smaller Moiré unit-cell size than predicted by the simulations is required to achieve robust coupling. Together with the dispersion-level agreement in Fig. 2, these cell-resolved spectra provide additional evidence of strong mode coupling in the lattice-mismatch Moiré cavity. Whereas Fig. 2 demonstrates agreement between calculations and measurements over the corresponding $k_x$ range, capturing both flat and dispersive features, Fig. 3 assesses coupling in the spatially extended array by comparing spectra from different unit cells. Consequently, our results indicate that the small-unit-cell lattice-mismatch Moiré configuration can support uniform collective resonances across the array.

**Mode-engineered Moiré nanolasers**

Figures 4a and 4b summarize the calculated characteristics of the lattice-mismatch Moiré cavity with a unit-cell size of 3.06 μm, as determined in Fig. 3. These include the Moiré unit-cell geometry and the calculated $|E|^2$ field profiles of the three flatband modes (hexapole, $p_x$, and $p_y$) (Fig. 4a), as well as the calculated band diagram plotted as a function of $k_x$ (Fig. 4b). The flatbands are spectrally isolated, thereby eliminating mode competition with other bands, which is advantageous for lasing. However, the near-frequency degeneracy of these three modes, which persists across different Moiré unit-cell sizes, complicates the realization of single-mode lasing. Reducing the hole size at the Moiré center enables single-mode operation: calculations for the modified structure, in which the central hole is removed, show that only the hexapole mode remains (Fig. 4c), leading to a single corresponding flatband in the band diagram (Fig. 4d). To conduct a more systematic investigation, we calculated the frequencies and $Q$ factors of the flatband modes as the central-hole scale decreased from 1.0 to 0.3 (Fig. 4e). The hexapole mode frequency remains nearly constant at ~193.2 THz, whereas the $p_x$ and $p_y$ mode frequencies decrease from 193.8 THz to 179.0 THz. Over the same central-hole scaling range, the $Q$ factor of the hexapole mode remains approximately $1.5 \times 10^4$, while the $Q$ factors of the $p_x$ and $p_y$ modes decrease substantially from 3180 to 445. These results indicate that removing the central hole does not significantly affect only the hexapole-mode flatband.

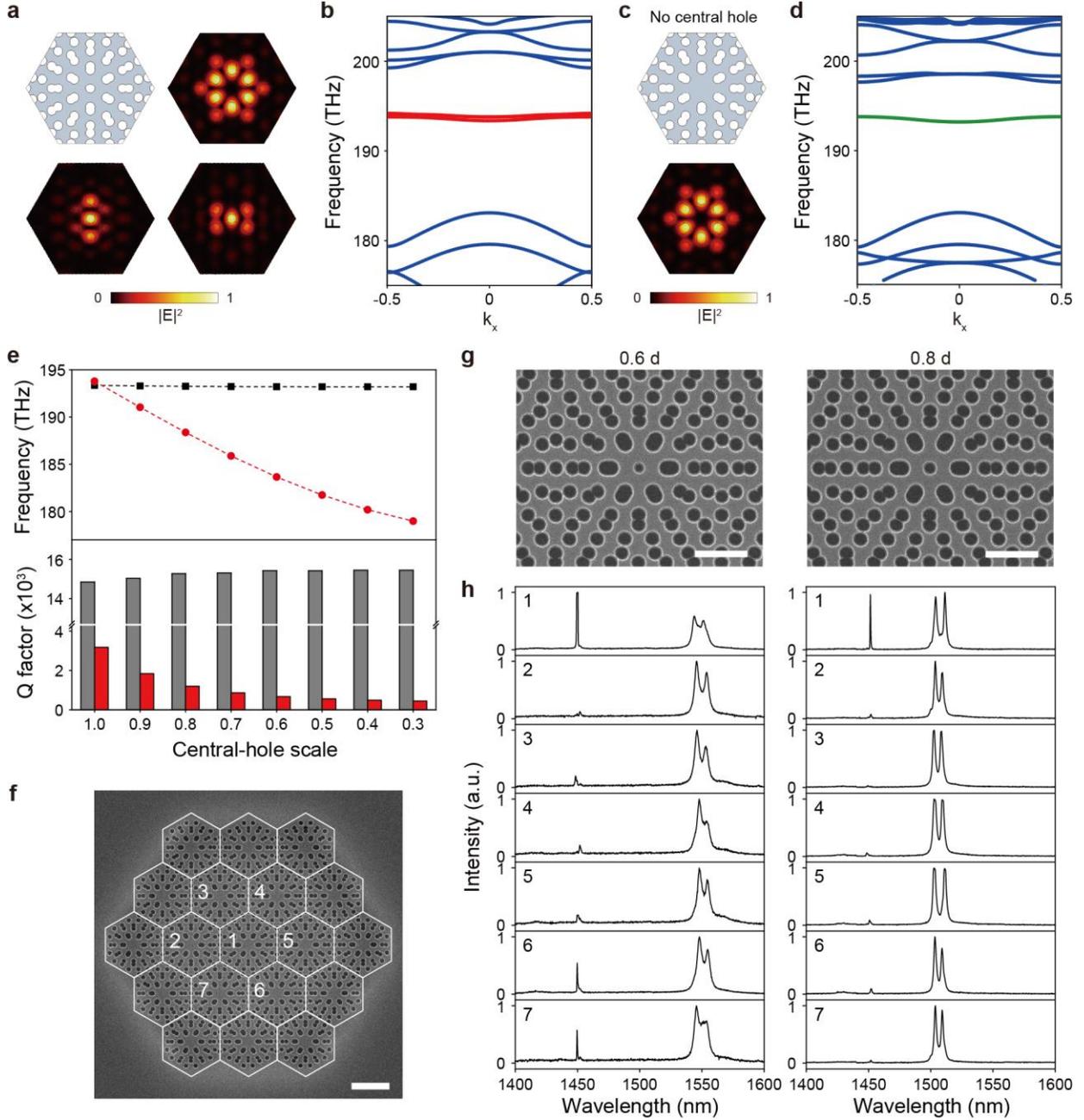

**Fig. 4. Flatband modes in lattice-mismatch Moiré cavities with modified central holes. a-d,** Calculated flatband modes and band diagrams for the original and modified lattice-mismatch Moiré cavities. **a,** Moiré unit cell of the original cavity (top left) and normalized electric-field intensity $|E|^2$ distributions of three flatband modes (hexapole, $p_x$, and $p_y$). **b,** Calculated band diagram of the original cavity in **a** along $k_x$, with flatbands highlighted in red. **c,** Moiré unit cell of the modified cavity with the central hole removed (top) and the normalized $|E|^2$ distribution of the corresponding hexapole flatband mode. **d,** Calculated band diagram of the modified cavity in **c**

along $k_x$, with the flatband highlighted in green. **e,** Calculated mode frequencies (top) and $Q$ factors (bottom) of the flatband modes as a function of the central-hole scale, ranging from 1.0 to 0.3 in steps of 0.1. Gray, hexapole mode; red, $p_x$ and $p_y$ modes. **f,** SEM image of a fabricated modified lattice-mismatch Moiré cavity. Seven core Moiré unit cells are labeled 1–7. Scale bar, 2 μm. **g-h,** Structural and spectral characterization of modified cavities with different central-hole scales. **g,** Magnified SEM images of cavities with central-hole sizes of 0.6 $d$ (left) and 0.8 $d$ (right), where $d$ is the original hole diameter. Scale bars, 1 μm. **h,** Measured PL spectra from the seven unit cells specified in **f** for cavities with 0.6 $d$ (left) and 0.8 $d$ (right). Each panel shows the spectrum from the correspondingly numbered unit cell over the wavelength range of 1400–1600 nm.

To examine this feature experimentally, we fabricated modified Moiré cavities with reduced central-hole sizes, as shown in the SEM image (Fig. 4f). Seven core Moiré unit cells are numbered from 1 to 7. We present the measured PL spectra from unit cells 1–7 for Moiré cavities with central-hole sizes of 0.6 $d$ (Fig. 4g, left) and 0.8 $d$ (Fig. 4g, right), where $d$ denotes the original hole diameter. The 0.6 $d$ cavity exhibits three peaks: a distinct sharp peak at 1449.6 nm and two broader, closely located peaks around 1550.5 nm (Fig. 4h, left). These peak wavelengths are identical throughout all seven unit cells (cells 1–7). Similarly, for the 0.8 $d$ cavity, three peaks are observed at consistent wavelengths across all seven unit cells, although the peak intensities vary from cell to cell (Fig. 4h, right). These observations highlight several key features. First, consistent with the Moiré cavities shown in Figs. 3a-c, the coupling of the flatband mode remains robust against variations in the central-hole sizes. Second, simple central-hole engineering enables the spectral separation of the flatband modes while preserving collective resonances; this separation is experimentally quantified by the larger spectral spacing observed for 0.6 $d$ ($\Delta\lambda = 100.9$ nm) compared with 0.8 $d$ ($\Delta\lambda = 55.5$ nm). Third, the experimental results show good agreement with the simulations. As shown in Fig. 4e, scaling the central hole separates the hexapole mode from the $p_x$ and $p_y$ modes, hence identifying a single high-$Q$ hexapole mode as a promising candidate for lasing.

We assessed the lasing characteristics of the modified lattice-mismatch Moiré cavities by analyzing the emission images, spectra, and light in–light out (L–L) curves (Fig. 5). Figure 5a shows the measured lasing images for four samples, arranged from top to bottom as 0 *d*, 0.4 *d*, 0.6 *d*, and 0.8 *d*, which represent the central-hole size. In the original Moiré design with a central-hole size of *d*, the emission spots exhibit uniform brightness across all seven unit cells. In contrast, as the central-hole size decreases, the brightness of the core unit cell increases in comparison to that of the six adjacent unit cells. In the measured spectrum of each cavity (Fig. 5b), we observe the following discernible features. First, the 0 *d* and 0.4 *d* cavities exhibit a single, distinct lasing peak at 1467.5 nm and 1464.7 nm, respectively. On the other hand, the 0.6 *d* and 0.8 *d* cavities display three peaks: a sharp peak at 1471.6 nm (0.6 *d*) and 1474.6 nm (0.8 *d*), together with two broader, closely located peaks at ~1563.5 nm (0.6 *d*) and ~1527.8 nm (0.8 *d*). This behavior agrees well with the calculated results presented in Fig. 4e. Second, when the central hole is significantly decreased (0 *d* and 0.4 *d*), only the hexapole mode remains within the gain spectrum, appearing as a single lasing mode. This suggests that the intense emission localized at the central unit cell originates from the hexapole mode. As the dipole modes become more prominent in the spectrum, bright emission emerges from the six adjacent unit cells. Third, the calculated normalized electric-field intensity profiles for each spectral peak show strong consistency with the experimental results (Fig. 5b, inset). The sharp peaks correspond to the hexapole mode, whereas the adjacent broad peaks (observed for 0.6 *d* and 0.8 *d*) are attributed to the $p_x$ and $p_y$ modes. We note that the mode intensity is highest in the central unit cell among the seven unit cells.

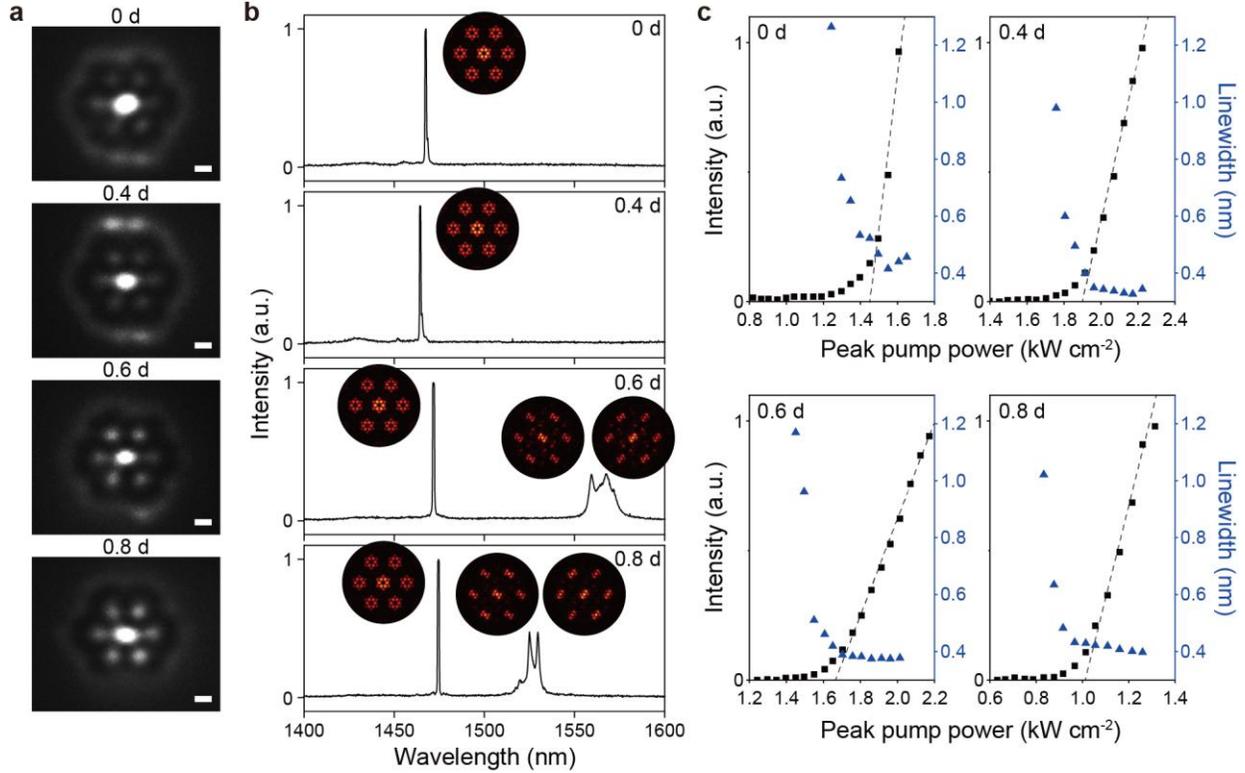

**Fig. 5. Lasing characteristics of modified lattice-mismatch Moiré cavities. a,** Measured real-space lasing images of cavities with four different central-hole sizes: 0 *d* (absence of the central hole), 0.4 *d*, 0.6 *d*, and 0.8 *d* (top to bottom). Scale bars, 2 μm. **b,** Measured PL spectra of cavities with central-hole sizes of 0 *d*, 0.4 *d*, 0.6 *d* and 0.8 *d* (top to bottom). Insets show the calculated normalized electric-field intensity $|E|^2$ distributions corresponding to the emission peaks. **c,** Measured L–L curves (black squares; left y-axis) and linewidths (blue triangles; right y-axis) for the emission peaks shown in **b** for central-hole sizes of 0 *d*, 0.4 *d*, 0.6 *d* and 0.8 *d* (top left to bottom right).

We also measured the emission intensity and linewidth of the lasing peaks shown in Fig. 5b as a function of pump power (Fig. 5c). Above threshold, the output power exhibits a clear superlinear increase, accompanied by a notable reduction in linewidth. The lasing thresholds for all four Moiré cavities are approximately 1–2 kW cm$^{-2}$, with a spectrometer-resolution-limited linewidth of ~0.4 nm. This lasing behavior is enabled by excitation of the targeted hexapole mode within the array, consistent with the Moiré periodicity requirements. Thus, we establish a direct design-to-demonstration approach for lasing in lattice-mismatch Moiré cavities. In particular, central-hole engineering validates effective mode separation and selective isolation of a single high-$Q$ hexapole resonance.

**Discussion**

We introduced, for the first time, a lattice-mismatch Moiré cavity and successfully developed a unique nanolaser enabled by strong flatband coupling. Optical simulations indicated that, unlike twist-angle Moiré cavities, this structure provides a nearly uniform flatband frequency shift, a substantial enhancement in the $Q$ factor relative to an isolated single-cell cavity, and a field distribution that extends across the unit-cell boundaries as the unit-cell size is reduced. These features collectively facilitate enhanced inter-cell interactions. The photonic band diagram of a small-unit-cell lattice-mismatch Moiré cavity was examined using PL emission from the sample, revealing a distinct flatband. Cell-resolved spectroscopy further confirmed that three flatband peaks appeared consistently across unit cells in the 3.06 μm unit-cell cavity. In contrast, the dominant peaks varied among unit cells in the 4.32 μm unit-cell cavity, indicating that the mode coupling was less robust than predicted by simulations. Furthermore, mode selection was achieved by reducing the central-hole size, thereby isolating the hexapole mode from the degenerate dipole modes, while preserving strong inter-cell coupling. Leveraging this feature, we successfully demonstrated hexapole flatband single-mode lasing in the lattice-mismatch Moiré cavity.

Validating distinctive phenomena predicted in Moiré physics within a nanophotonic platform is significant[2,13]. By implementing lattice-mismatch Moiré cavities and systematically adjusting the lattice-constant difference in conjunction with the Moiré unit-cell size, we provide an experimental approach to maintaining and verifying strong inter-cell coupling beyond what can be inferred solely from band-structure measurements. Furthermore, we consistently position the flatband within a targeted spectral range and inside the photonic bandgap. By extending Moiré photonics from conventional twist-angle configurations to the lattice-mismatch regime, our results demonstrate that systematic manipulation of the relative lattice parameters of the two constituent lattices offers a practical and versatile design strategy for optimized Moiré nanolasers and flatband nanophotonic devices.

## Methods

**FEM simulations.** Three-dimensional finite element method (FEM) simulations were performed using the Wave Optics Module of COMSOL Multiphysics to calculate optical resonant modes and band structures. Floquet periodic boundary conditions were employed for coupled-mode calculations in periodic structures, while scattering boundary conditions were used for single-unit-cell calculations. Band structures were calculated under periodic boundary conditions by varying the Bloch wavevector. The refractive indices of the InGaAsP slab and air were set to 3.33 and 1.0, respectively.

**Device fabrication.** The samples were fabricated on an epitaxial wafer consisting of a 257-nm-thick InGaAsP slab containing three quantum wells, a 1-μm-thick InP sacrificial layer, a 100-nm-thick InGaAs etch stop layer, and an InP substrate. The quantum wells were designed for a central emission wavelength of ~1.5 μm. A poly(methyl methacrylate) (PMMA) layer was spin-coated on the wafer, and electron-beam lithography at 30 keV was used to define the hole patterns. The patterned PMMA served as an etch mask, and the air holes were etched into the InGaAsP slab by inductively coupled plasma reactive ion etching (ICP-RIE). After etching, the PMMA was removed using $O_2$ plasma, and the InP sacrificial layer was selectively wet-etched using a 3:1 $HCl:H_2O$ solution, resulting in a free-standing InGaAsP photonic crystal slab.

**Optical measurements.** A 980-nm pulsed laser diode (1.0% duty cycle and 1-MHz repetition rate) was used to optically pump the samples with a spot size of ~29.2 μm. The emission from the lattice-mismatch Moiré cavities was collected using ×50 and ×100 objective lenses with numerical apertures of 0.42 (378-805-3, Mitutoyo) and 0.85 (LCPLN100XIR, Olympus), respectively. The PL signal was directed to either an InGaAs IR camera (PA1280F30NCLd, Ozray) or an IR 2D array detector (IsoPlane-320 and NRV-640, Princeton Instruments). The spectrometer resolution was ~0.4 nm. The measured L–L curves were plotted as a function of the peak pump power (Fig.

5c). For band-diagram measurements (Fig. 2), the Fourier-plane (*k*-space) image was spatially filtered using a vertical slit and then spectrally dispersed by a spectrometer. The slit was positioned to select a specific Bloch wavevector (in-plane wavevector) axis, enabling band-structure measurements along the chosen direction in reciprocal space. The measurable *k*-range in a single acquisition was determined by the slit width and the numerical aperture of the collection optics, yielding 2.63 $\mu m^{-1}$ for the ×50 objective lens and 5.44 $\mu m^{-1}$ for the ×100 objective lens.

**Data availability.** The data that support the findings of this study are available from the corresponding authors upon request.

**Code availability.** The codes used in this work are available from the corresponding authors upon request.

**Acknowledgements**

This work was supported by the Samsung Science and Technology Foundation (project no. SSTF-BA2401-02) and the National Research Foundation of Korea (NRF) grant funded by the Korean government (MSIT) (2021R1A2C3006781).

**Author contributions**

D.K. and H.-G.P. conceived the idea. D.K. designed the cavities and conducted the simulations. D.K. and C.S. performed the optical measurements. D.K., C.S., G.-W.L., and Y.-S.N. fabricated the samples. D.K., C.S., C.K., J.-K.Y., H.J., and H.-G.P. analyzed the experimental and theoretical data. All authors contributed to writing and editing the manuscript.

**Competing interests**

The authors declare no competing interests.

**Correspondence and requests for materials** should be addressed to H.J. or H.-G.P.

# Supplementary Information

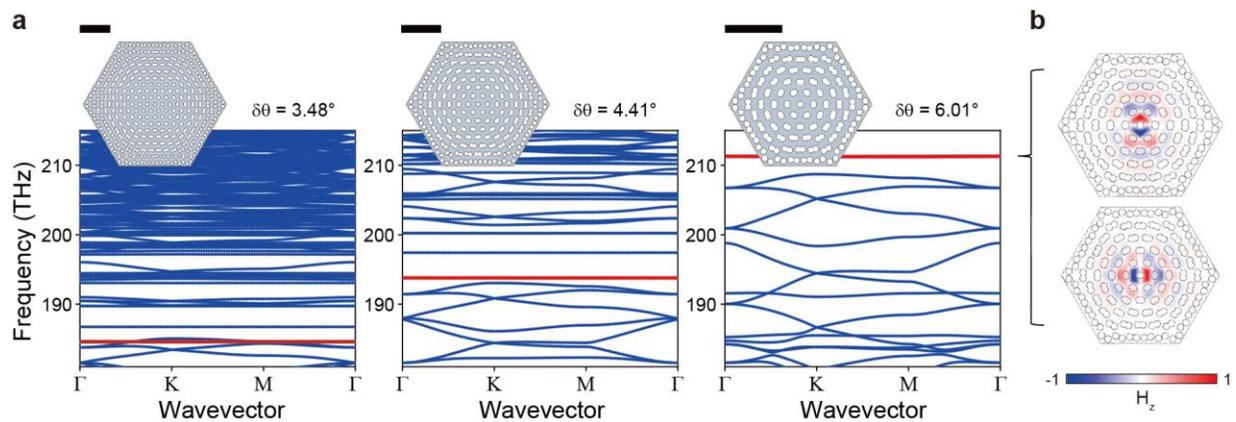

**Supplementary Fig. 1. Simulations of twist-angle Moiré cavities. a,** Calculated band diagrams of twist-angle Moiré cavities with $\delta\theta$ = 3.48°, 4.41°, and 6.01° (from left to right) in the frequency range of 181–215 THz, where the flatbands are highlighted in red. The corresponding Moiré unit-cell structures are shown as insets. Inset scale bars, 2 μm. **b,** Calculated $H_z$ field distributions for two flatband modes at $\delta\theta$ = 6.01°. Two degenerate dipole modes: $p_x$ (top) and $p_y$ (bottom).

**a** Twist-angle Moiré

| $\delta\theta$ (°) | Unit-cell size (μm) | Triangular lattice $a$ (nm) | Flatband frequency (THz) | $\Delta f$ (THz) |
|---|---|---|---|---|
| 3.48 | 8.16 | 496 | 184.61 | - |
| 4.41 | 6.24 | 480 | 193.78 | 9.17 |
| 6.01 | 4.32 | 453 | 211.28 | 26.67 |

| Single-cell $Q_{single}$ | Periodic Q | $Q/Q_{single}$ |
|---|---|---|
| 64916 | 190040 | 2.93 |
| 14183 | 134900 | 9.51 |
| 2394.7 | 63991 | 26.72 |

**b** Lattice-mismatch Moiré

| $\delta a$ (nm) | Unit-cell size (μm) | Triangular lattice $a_1$ (nm) | Triangular lattice $a_2$ (nm) | Flatband frequency (THz) | $\Delta f$ (THz) |
|---|---|---|---|---|---|
| 30 | 8.16 | 480 | 510 | 188.61 | - |
| 40 | 6.24 | 480 | 520 | 189.46 | 0.85 |
| 60 | 4.32 | 480 | 540 | 190.50 | 1.89 |

| Single-cell $Q_{single}$ | Periodic Q | $Q/Q_{single}$ |
|---|---|---|
| 31659 | 196820 | 6.22 |
| 3444.7 | 145170 | 42.14 |
| 383.93 | 114630 | 298.57 |

**Supplementary Fig. 2. Comparison between twist-angle and lattice-mismatch Moiré cavities. a-b,** Design parameters and calculated properties for the twist-angle Moiré (**a**) and lattice-mismatch Moiré cavities (**b**).

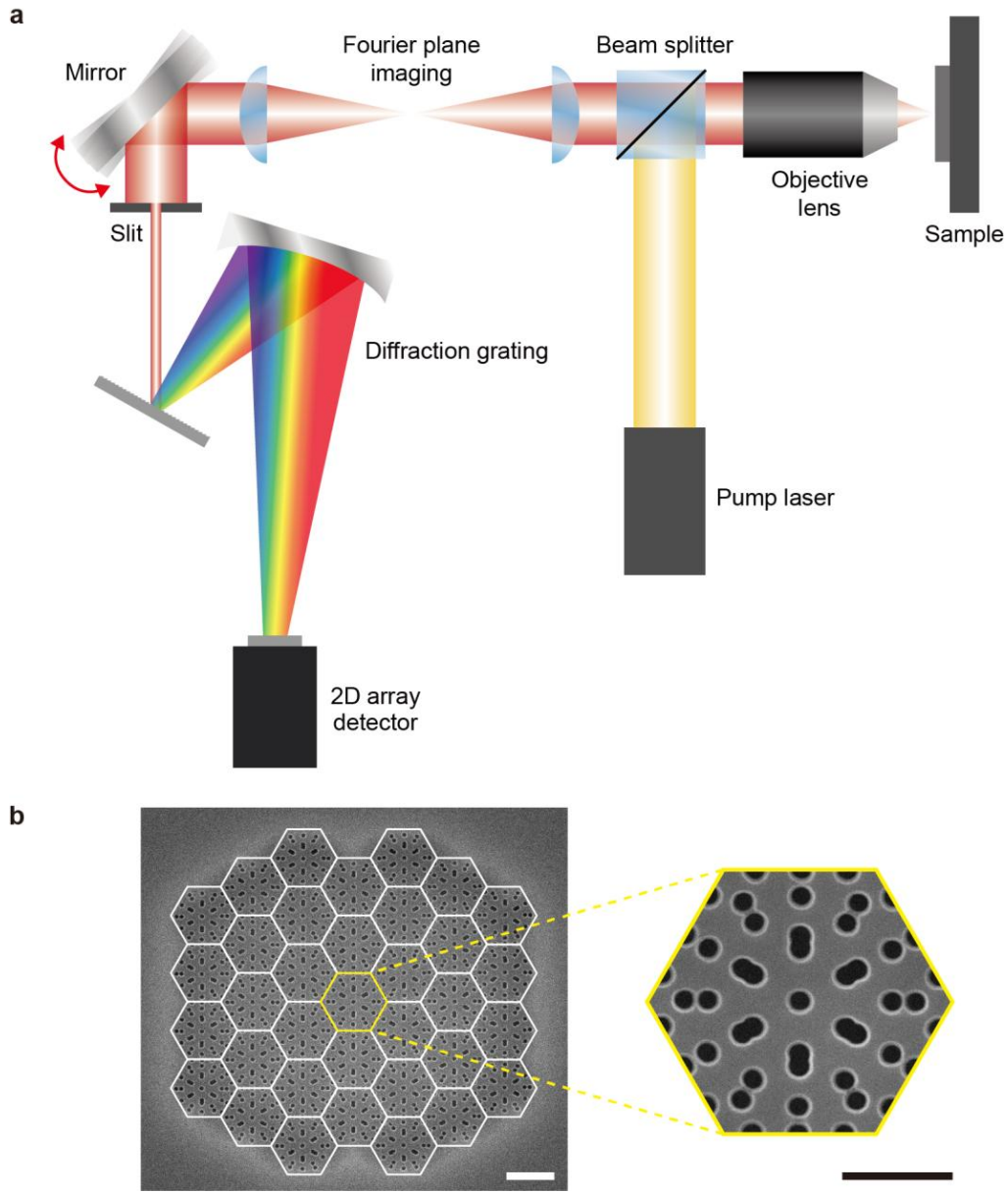

**Supplementary Fig. 3. a,** Schematic of the band measurement setup. **b,** SEM images of a lattice-mismatch Moiré cavity used for band measurements. The right SEM image (in yellow) shows a magnified central unit cell. Scale bars, 2 μm (left) and 1 μm (right).

# Supplementary Note

## 1. Construction of lattice-mismatch Moiré superlattice

This section presents the construction procedure for lattice-mismatch Moiré superlattices formed by superimposing two triangular lattices with different lattice constants. For a fixed $a_1 = 480$ nm, Fig. R1 lists the possible $a_2$ values that generate a lattice-mismatch Moiré pattern, along with the corresponding unit-cell sizes.

| $\delta a$ (nm) | Triangular lattice $a_2$ (nm) | $a_1 : a_2$ | Unit-cell size (µm) | $\delta a$ (nm) | Triangular lattice $a_2$ (nm) | $a_1 : a_2$ | Unit-cell size (µm) |
|---|---|---|---|---|---|---|---|
| 240 | 720 | 2 : 3 | 1.44 | 40 | 520 | 12 : 13 | 6.24 |
| 160 | 640 | 3 : 4 | 1.92 | 36.9 | 516.9 | 13 : 14 | 6.72 |
| 120 | 600 | 4 : 5 | 2.40 | 34.3 | 514.3 | 14 : 15 | 7.20 |
| 96 | 576 | 5 : 6 | 2.88 | 32 | 512 | 15 : 16 | 7.68 |
| 80 | 560 | 6 : 7 | 3.36 | 30 | 510 | 16 : 17 | 8.16 |
| 68.6 | 548.6 | 7 : 8 | 3.84 | 28.2 | 508.2 | 17 : 18 | 8.64 |
| 60 | 540 | 8 : 9 | 4.32 | 26.7 | 506.7 | 18 : 19 | 9.12 |
| 53.3 | 533.3 | 9 : 10 | 4.80 | 25.3 | 505.3 | 19 : 20 | 9.60 |
| 48 | 528 | 10 : 11 | 5.28 | 24 | 504 | 20 : 21 | 10.08 |
| 43.6 | 523.6 | 11 : 12 | 5.76 | ⋮ | ⋮ | ⋮ | ⋮ |

**Fig. R1.** List of triangular-lattice configurations for constructing a lattice-mismatch Moiré superlattice with a fixed lattice constant $a_1 = 480$ nm. For each commensurate ratio $a_1 : a_2$, the table reports the corresponding lattice constant $a_2$, the lattice-constant difference $\delta a = |a_1 - a_2|$, and the resulting Moiré unit-cell size. The configurations used in the main text are highlighted in red.

As shown in Fig. R1, the Moiré unit-cell size is determined by the least common multiple (LCM) of $a_1$ and $a_2$. Here, we choose successive integer ratio $a_1 : a_2 = n : (n+1)$ as our commensurate condition. This condition enables the formation of a lattice-mismatch Moiré cavity that supports flatband modes.

A significant advantage of the lattice-mismatch Moiré pattern is the tunability of the unit-cell size. Once the commensurate condition $a_1: a_2 = n : (n + 1)$ is satisfied, the Moiré unit-cell size can be increased arbitrarily by choosing larger $n$, in principle without an upper bound. Unit-cell sizes intermediate to the discrete values listed in the table can also be achieved by uniformly scaling both $a_1$ and $a_2$. For example, the 3.06 μm unit cell used in the main text corresponds to a ×1.0625 scaling of the 2.88 μm unit cell associated with $a_1: a_2 = 5 : 6$. Maintaining the same ratio, $a_1$ and $a_2$ become 510 nm and 612 nm, respectively ($\delta a$ = 102 nm).

## 2. Analysis of photonic flatbands using a tight-binding model

The origin of the photonic flatband can be explained using a simple one-dimensional tight-binding model[S1,S2]. We consider a one-dimensional bipartite stub lattice (Fig. R2), where circles denote lattice sites with field amplitudes $\psi_n^j$, and lines indicate nonzero inter-site hopping terms. In this geometry, the $b$ sites form the majority sublattice, so hopping between the minority sites necessarily occurs via the majority sites, with no direct minority-minority coupling.

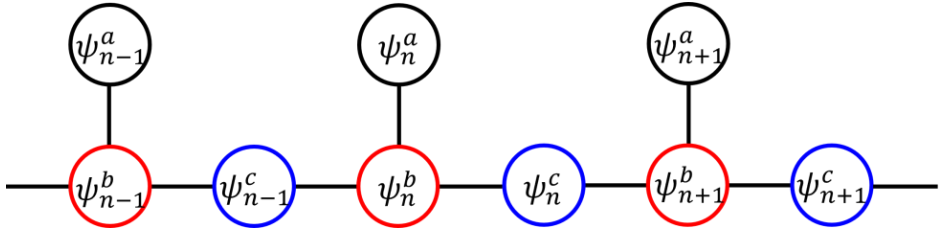

**Fig. R2.** 1D bipartite stub lattice with three site types: one majority site ($b$) and two minority sites ($a$ and $c$).

The tight-binding equations for this model are given by

$$\delta\omega\psi_n^a = J_1\psi_n^b \tag{1}$$

$$\delta\omega\psi_n^b = J_1\psi_n^a + J_2\psi_{n-1}^c + J_3\psi_n^c \tag{2}$$

$$\delta\omega\psi_n^c = J_2\psi_{n+1}^b + J_3\psi_n^b \tag{3}$$

where $\delta\omega$ is the detuning relative to a reference frequency, $J_1$, $J_2$ and $J_3$ are inter-cell coupling constants, and $\psi_n^a$, $\psi_n^b$, and $\psi_n^c$ are amplitudes at each site in the $n$th unit cell. Solving these coupled equations yields the energy eigenvalues,

$$\delta\omega = 0, \pm\sqrt{J_1^2 + J_2^2 + J_3^2 + 2J_2J_3\cos ka} \tag{4}$$

where $k$ is the wavevector and $a$ is the lattice period. $\delta\omega = 0$ implies a $k$-independent eigenfrequency, i.e., a flatband with a constant frequency.

We use this simple model to provide an intuitive explanation for the emergence of a flatband in our lattice-mismatch Moiré structure. Specifically, we conceptually "close" the lattice by bending it and connecting its two ends: a given $a$ site is surrounded by an alternating sequence of $b$ and $c$ sites. This closed-geometry picture can be directly mapped onto the lattice-mismatch Moiré structure, as illustrated in Fig. R3.

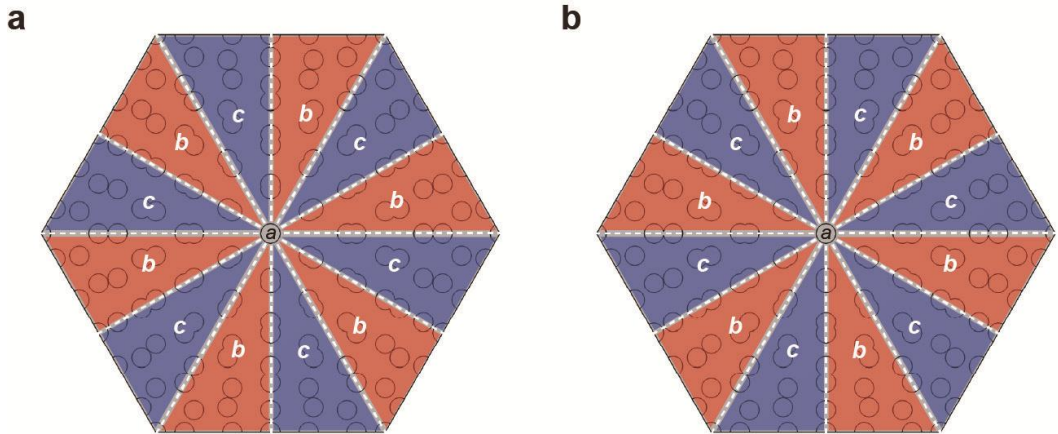

**Fig. R3.** 1D bipartite stub lattice mapped onto a lattice-mismatch Moiré structure, comprising three types of sites ($a$, $b$, and $c$). Panels **a** and **b** show two distinct mappings corresponding to different choices of the majority site.

Because the lattice-mismatch Moiré unit cell (and the overall lattice) possesses $C_6$ symmetry, it can be viewed as a ring-like structure in which the $b$ and $c$ sites alternate and repeat six times around a central $a$ site. Depending on which site type is treated as the majority, two distinct mappings are possible (Figs. R3a and R3b). These two mappings correspond to different choices of the reference frequency (i.e., different on-site frequency offsets depending on which sites are treated as the majority), implying the existence of two flatband solutions.

As discussed in the main text, the lattice-mismatch Moiré cavity supports three flatband modes (hexapole, $p_x$, and $p_y$). The two mappings in Fig. R3 account for two of these solutions, while the remaining one can be understood using the picture shown in Fig. R4.

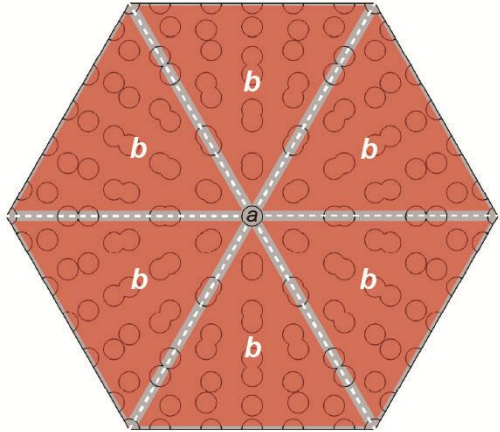

**Fig. R4.** 1D bipartite stub lattice mapped onto a lattice-mismatch Moiré structure comprising two site types (*a* and *b*).

This configuration can likewise be understood as a "bent" realization of a one-dimensional bipartite stub lattice composed of two sites (Fig. R5).

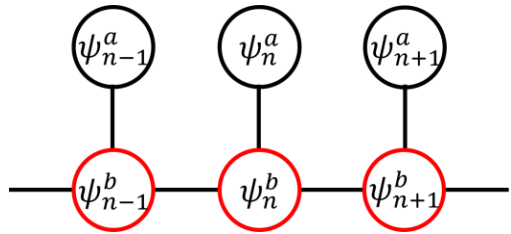

**Fig. R5.** 1D bipartite stub lattice with two site types: a majority site (*b*) and a minority site (*a*).

The tight-binding equations for this model are given by

$$\delta\omega\psi_n^a = J_1\psi_n^b \tag{5}$$

$$\delta\omega\psi_n^b = J_1\psi_n^a + J_2\psi_{n-1}^b + J_2\psi_{n+1}^b \tag{6}$$

where $\delta\omega$ is the detuning relative to a reference frequency, $J_1$ and $J_2$ are inter-cell coupling constants, and $\psi_n^a$ and $\psi_n^b$ are amplitudes at each site in the $n$th unit cell. Solving these coupled equations yields the energy eigenvalues,

$$\delta\omega = J_2 \cos ka \pm \sqrt{(J_2 \cos ka)^2 + J_1^2} \tag{7}$$

If $J_2$ is much larger than $J_1$ (i.e., $J_2 \gg J_1$), the following approximation holds:

$$\delta\omega \cong 0, 2J_2 \cos ka \tag{8}$$

This implies a flatband with a constant frequency.

This approximation is well justified for our lattice-mismatch Moiré structure. Due to the $C_6$ symmetry of the lattice-mismatch Moiré unit cell (and the overall lattice), the coupling between $b$ sites, $J_2$, is expected to be much larger than the coupling between the $a$ and $b$ sites, $J_1$. Therefore, Eqs. (4) and (8) together explain the mechanism by which the three flatbands form in the lattice-mismatch Moiré structure.

## Supplementary References